**Laser Welfare: First Steps in Econodynamic Engineering**

**Geoff Willis**

gmw@intonet.co.uk

Abstract

The paper starts with a brief review of present understanding of income distributions; especially with regard to recent work in the field of econophysics that draws parallels between income, wealth and energy distributions.

Examples of alternative energy distributions found in physical systems are discussed, and how they could be used to construct economic models that might allow alternative overall distributions of wealth and income in society.

These ideas are further expanded and a more detailed scheme for welfare assistance is proposed that might be used to improve the incomes of the poorest in a more efficient way than traditional welfare schemes.

Finally, and unusually for a paper on economics; an experiment is proposed that could be used to either support or disprove the ideas discussed in the rest of the paper.

Contents

1. Background: Income Distribution and Statistical Physics

Since the work of Pareto (1) as long ago as 1897, it has been known that distributions of wealth or income have appeared to be log normal distributions that strictly follow power law decays. These distributions have been observed across a wide variety of different economies over long periods of time. Traditionally this has been an economic puzzle; as intuitively different income distributions would be expected in differently structured economies.

In recent years, the study of income distributions has gone through a small renaissance with new interest in the field shown by physicists with an interest in economics.

Wataru Souma has carried out extensive and detailed research into large data sets, primarily from Japanese income data (2,3). He has demonstrated with great clarity that income distributions consistently have a power tail decay at high income levels. He also demonstrates that the log-normal distribution gives a good fit to Japanese income data at lower income levels. The distributions for some of the years analysed show a smooth transition from the low income, log-normal curve to the high income, power tail distribution. However in some years there is a clear discontinuity between the two distributions.

Victor Yakovenko, Adrian Dragulescu and A Christian Silva (4,5) have also shown the presence of a Pareto power law tail in analysis of US, UK and Japanese income tax data. They have further proposed a Boltzmann-Gibbs exponential mid-section to the same data sets, and demonstrated that the parameters controlling the power tail are strongly connected to stock-market indices.

Independently, Jean-Phillipe Bouchaud & Marc Mezard (6), and Sorin Solomon & Peter Richmond (7a, 7b) have proposed stochastic approaches using complexity theory to study income distribution. These have been very successful in explaining the origins of the power tails observed in income distributions, the first plausible theoretical frameworks to be proposed in the history of the study of income distribution. These frameworks have been less convincing in explaining the apparent log-normal distribution found at lower incomes. (Solomon & Richmond have also introduced the phrase 'econodynamics' to describe the similarities with a thermodynamic approach).

Juergen Mimkes and the author (8) have done a detailed analysis of UK and US data that shows a clear Boltzmann relationship ($x.\exp(-x)$) for waged income data. Along with Th Fruend, they have also proposed a relatively straightforward maximum entropy equilibrium approach (9) to explain the source of these Boltzmann distributions. (The author first proposed that income/wealth distributions are Boltzmann distributions in 1993 and hinted at ways of modifying these distributions at this time (10).) However this approach gives no power tail, and so does not explain the distributions seen at higher incomes.

As as example; Figures 1a & 1b demonstrate the close correlation that can be seen between actual economic data and Boltzmann distributions in UK and US income data.

Independently Duncan Foley (11), Levy, Solomon, and others (review (12)) have proposed a maximum entropy, agent based, approach that produces a distribution that closely mimics the log normal shape.

Foley (13) has expanded these ideas and used the fact of the effective quantisation of employment and welfare to explain the persistence of mass unemployment under certain circumstances where a maximum entropy approach leads to a failure in market clearing that is not seen in a Walrasian approach.

Finally, Arnab Chatterjee, Bikas Chakrabarti & S. S. Manna (14), Arnab Das & Sudhakar Yarlagadda (15) and have all used agent based approaches to income distribution based on gas-like market models. When saving propensity is fixed in these models, the outcome is a Gibbs like distribution. When the agents have a random saving propensity in the models, the distributions have power tails. Similar results have also been produced by Salete Pianegonda and Jose Roberto Iglesias (16).

Although this has not been formally accepted in the wider economics community; this paper takes as a starting point the assumption that income and wealth are indeed distributed on an 'econodynamic' basis, and that the overwhelming majority of income is distributed as a Boltzmann-Gibbs function as a result of maximum entropy considerations.

For physicists and others that are familiar with a maximum entropy approach it is not then a surprise that the same distributions of income are seen in widely different economic systems. From a statistical mechanical point of view; as the number of participants in a system increases, the underlying mechanisms of exchange (whether this is of energy or income) become irrelevant; and the resulting distribution is simply the one that is statistically most likely.

This has very important consequences, both for the effects that such distributions have on human life, and the ways that human beings can affect these distributions.

Firstly it is worth considering the appropriateness of the Boltzmann distribution as a method for sharing wealth amongst humanity. Most human abilities are found to be distributed on the basis of a normal distribution as shown in Fig. 2. The tails of this distribution decline to zero rapidly, and the mode usually has a large offset from zero when describing human qualities. In such a distribution the mean, median and mode averages coincide very closely.

The result is that for most human skills the variation in ability between the top decile and the bottom decile is only of the order of a factor of two or three or so, very rarely by factors of ten or more.

The above is not true of course for learnt skills, it is however generally true of the ability to learn these skills. Given these ranges of human abilities it is possible to construct an argument that a "fair" economic distribution would be one similar to an offset normal distribution.

The Boltzmann distribution however is markedly different in two important respects. Firstly it is skewed; the mode average is considerably below the mean average, with the median somewhere between these two. Secondly it has a long tail with significant numbers of extreme events populated at levels considerably above either the median or mean averages. In a Boltzmann distribution (see Fig. 3.) the bottom decile lies close to the zero axis and has wealth significantly less than the average wealth. The top decile has wealth considerably in excess of the average wealth. Two other things can be noted with regard to the distribution. Firstly, that the displacement between the mean and the median results in a significant majority of individuals having less than the mean value of wealth. Secondly there is automatically a portion, roughly 15% of all individuals, who are permanently below half the mean average income, that is the normally defined poverty level.

It is possible from the above to construct an argument that the Boltzmann distribution is not a "fair" way for wealth or income to be distributed in a society. Certainly social democrats, socialists and communists have constructed such arguments and have offered differing solutions to solving this perceived problem.

In communist states strict, and active, microeconomic control was the normal way of attempting to prevent large discrepancies in wealth. In democratic countries this has generally been avoided, because of the stunting effects on economic growth. Instead these countries have instituted massive systems of taxation and welfare in an attempt to transfer wealth from the rich to the poor. Meanwhile trade unions and professional societies also attempt to modify wealth distributions for there own members.

From an econophysic point of view the above methods of attempting to influence wealth distribution are deeply flawed. In a system of a large number of freely interacting particles the Boltzmann distribution is inevitable and methods of exchange, even ones such as tax and welfare, are largely irrelevant.

From an econodynamic perspective, an approach that does make some sense is that of the trade unionists and professional societies. By tying together the interests of thousands, or even millions, of individuals their members are no longer "freely interacting" and are able to release themselves from the power of entropy to a limited extent. (Monopolistic companies attempt to subvert entropy by similar means).

Traditional methods of taxation and welfare seem to have much less justification. It is common experience that such transfers give little long term benefit to the poor. Transfers need to be massive and continuous to be effective, and there is a wealth of data to suggest that many welfare programmes result in the giving of benefit to those of medium income, rather than to the poor. This is of course exactly what an econodynamic analysis would predict.

Given the power of entropy to force the overall distribution regardless of different sorts of microeconomic interactions, it would initially seem that attempts to modify income distribution will be futile. This is not necessarily the case.

One approach to modifying distributions is to look at market failures, such as those discussed by Foley, from an econodynamic point of view, and use economic policy to ensure that these regions of failure are not allowed to occur. So, for example, the quantisation effects that Foley discusses might be avoided by such measures as job subsidies and negative income tax.

An alternative approach is to look at analogies from other physical systems which could be used as alternative economic models. I intend to follow this approach in the next two sections of this paper.

2. Alternative Distributions

The distinctive skewed shape of the Boltzmann distribution is a result of the particular boundary conditions found in most energetic systems. In an ideal gas the positive energy of any individual molecule is effectively unlimited; the molecule can go as fast as it wants. There is however a very clear boundary condition in the other direction, it is impossible for any molecule to have negative energy, once it is stopped it can not go any slower. It is this boundary condition that forces the skewness in the Boltzmann distribution. For a few of the molecules to have a lot of energy it is not possible for a few to have a lot of negative energy. Instead a summation over all possible assemblies dictates that a lot of molecules must have a little energy to compensate for the few with a lot of energy.

(In economic systems this zero boundary condition is due to the difficulty any human being has in maintaining significant long term values of negative wealth; a growing problem with the increasing levels of communication between credit agencies.)

There are a small number of systems that do not show the typical Boltzmann distribution of energy, the most obvious of which is the laser.

A non-lasing material has closely spaced energy gaps. The exchange of photons between molecules can result in free interchange of energy states, with one molecule increasing in energy and the other decreasing; as shown in Fig. 4.

A lasing material typically has a closely spaced band of lower energy levels with a large gap between this band the next one above. If a molecule is already at the top of this lower band (at E4) it is unable to go to the next energy level up because there is no other molecule available in bands E1-E4 that can make an equal large jump down. In Fig. 5. $\Delta E_{4-5}$ is greater than $\Delta E_{1-4}$, so no molecule can jump from E4 to E5, because their is no matching drop available to keep total energies balanced. If the material is kept isolated, this lower band can be given a very high occupation of energy levels "pumped" by an external source. Again, if kept isolated; this "inverted" distribution can be maintained for a significant time. Such an inverted distribution is not inherently thermodynamically stable; which is part of the reason that the release of energy is so intense when a laser is allowed to interact with its external environment.

The reason for the inversion of the distribution is the existence of an effective upper limit on the distribution. It is possible that such an approach could be used in an economic system.

Given a hypothetical isolated economic population N with total Wealth W and average wealth w= W/N, let us assume that a law is introduced that dictates that any individual that has more than double the average wealth is committing a criminal offence and is jailed. So the range of assemblies over which the total possible distributions is to be calculated is now limited at 2w instead of infinity. Any distribution that has a person with wealth greater than 2w must be discarded from the total of assemblies.

By symmetry this would result in bell shape running from zero at zero, to zero at 2w, and having a maximum at w; Something like Fig. 6, and similar to a "wide" normal distribution offset from zero to w.

Such a distribution would move a significant group of people out of the lower wealth levels and could be perceived as being more fair in its overall sharing of wealth.

It is possible to go further; if the maximum wealth were set at 1.5 times the average wealth say, then a laser like inverted distribution of the form shown in Fig. 7. would result. Statistically the occupation of the lower levels would be very small indeed.

While these ideas are theoretically sound they have very fundamental flaws as practical ways of running modern economies.

The first obvious problem is that of isolation. If a maximum wealth (or maximum income) law was introduced in a typical western economy then the individuals affected by the law would simply move themselves or their excess wealth to another economy without such a law. To be effective such a maximum wealth law must either apply to a nation that is isolated with strict controls on both emigration and financial transfers, or the law would have to be applied simultaneously to all interacting economies without the exemption of a single offshore tax haven.

The second obvious problem is the process of actually moving from a free Boltzmann distribution to a "capped" distribution. This would involve a substantial reduction in wealth for a significant (and influential) proportion of the population. Perhaps more importantly it might create a perceived loss of opportunity for a much larger portion of the population, and would almost certainly be seen as an infringement of liberty in most societies.

(The preceding discussions have largely ignore the power tail that is known to be present in income data. The presence of the tail increases the validity of the arguments in matters such as the perceived 'unfairness' and also in the effectiveness of proposed solutions to reduce this 'unfairness'.)

In the next section I would like to further refine these concepts to produce much smaller "econodynamically isolated" economies that might be more practically applied in the augmenting of the income of the poorest members of society.

3. Creating Local Isolated Systems - Laser Welfare

In this section a welfare system is proposed that would operate within an economy but be isolated from the economy, apart from the subsidies needed to keep the isolated system functioning.

The system to be created will consist of one thousand unemployed people. These people are assumed to be "priced out of the market" their existing skills are not productive enough for them to be attractive to an employer in the open market. It is intended to provide subsidies to these people to help them into jobs. These subsidies will be equivalent to the "pumping" that takes place in a laser.

To build the system a number of items are needed. The first is a separate system to represent financial wealth. In this system these will be referred to as "coupons". Each coupon (C1) Will be redeemable from the government for $1 cash.

The second item needed is a way of limiting the number of coupons an individual is allowed to earn. This is achieved by issuing each person in the scheme an allowance book. Each week the beneficiary will be allowed to claim cash from the government against any coupons they have gained that week up to a maximum of each week's allowance. Allowances are strictly non-transferable, whilst coupons are freely tradable.

The government uses subsidies and the free market to keep the system circulating. It is assumed that the beneficiaries are only 50% of the efficiency of a typical person employed on low wages. It is also assumed that the government wants each beneficiary to earn around $100/week; this is deemed to be sufficient to meet their basic needs. Suppose there are 1000 beneficiaries in the scheme, then each beneficiary is given an allowance book that allows the beneficiary to cash in up to, but no more than, 120 coupons (C120) each week. Each week the total of available allowances will be C120 000.

However each week the government will only release C100 000. These coupons will be released by auctioning them to a number of registered employers taking place in the subsidized labour scheme. The employers will purchase the coupons from the government each week for cash. The employers will then exchange them with the beneficiaries in exchange for their (inefficient) labour.

This may seem a very complex way of getting money into the hands of unemployed people, but it does have some positive effects. A micro economy has been created in which different employers compete to buy coupons from the government, and different beneficiaries compete with each other. However the competition for the beneficiaries is not so fierce. A closed system has been created in which average occupancy is 83% (100/120), an inverted distribution will therefore result, and the number of beneficiaries earning less than say $80 will be very small.

Initially the price paid by the employers to the government would probably be very low, $0.1/Coupon or something of this order, but competition should drive this price up to around $0.5/Coupon, as we assumed our beneficiaries were 50% efficient (this ignores bureaucratic costs).

This is very useful for the government as they are now only paying out a net $0.5 for each $1 that reaches the pocket of the beneficiary.

However the process should not stop there. With competition in place it is to the advantage of both the employer and the beneficiary to improve the efficiency of the beneficiary. The employer that is able to make the best use of the employee is the one that will make the most profit at a certain auction price for coupons. The beneficiary that can increase the value of his skills to his employer is the one that is likely to earn closer to 120 coupons rather than only end up with 80 coupons. Healthy competition has advantages for both parties.

(Section 7. of this paper gives a more longwinded demenstration of how such a system might work.)

Clearly the above is very simplistic and will need substantial research and development before it could be worked into a real life welfare program. Like any welfare or taxation program there will be opportunities for fraud; ways in which human beings can breakdown the barrier between the two "isolated" systems.

However the main point that is being made is that a knowledge of the principles of econophysics / econodynamics may be potentially used to create alternative and effective financial systems.

4. Experimental Verification

In this section a possible experiment is discussed that would either support or disprove the ideas discussed in the proceeding sections.

The aim is to create isolated economic communities that would be allowed to interact in identical ways, except that in the first experiment (A) there would be no restriction on individual wealth, while experiment B would have such a restriction.

The proposed format would be to use a university hall of residence during holiday time. The two experiments would take place consecutively in the same hall to give identical conditions. Each experiment would include a minimum of 50 students and would take between one and two weeks.

Initially students would be provided with dormitory accommodation with all men in one communal room and all women in another, the mattresses in these two dormitories would be of the minimum comfort. Students would be provided with free minimum daily rations that would consist of rice, onions, tomatoes, beans, salt and cooking oil only.

During the day 50 exercise bicycles would be provided that the students could use to earn credits during the day.

Each evening there would be a sale / auction for various commodities.

The following "tangible" assets would be sold at fixed rates advertised in advance.

- various items of "luxury" food such as meat, cheese, eggs, bread, butter, flour, sugar, coffee, tea, beer, wine, etc.
- use of thicker mattresses for one night only.

- use of newspapers, books and magazines for one night only (and only one copy of each newspaper or book would be available).
- radios, cd players and cd's (again numbers of each item would be limited).

The following "rental" items would be auctioned.

- hourly access to a limited number of television rooms for the rest of the evening.
- use of a private single or double room for one night only.
- hourly access to each exercise bike for the following day.

All the exchanges would be done on a computer based system with a record of the actual commodity exchanged. Following the auction, students would be allowed to exchange commodities at rates agreed by themselves, but again all exchanges would be recorded.

On the first day each student would be given a free and equal slot of access to the exercise bikes. But from the first evening this access (access to the only means of production) would be the first item in the auction, with any individual allowed to bid for as many hours as they want and resell the access later.

Following this the other rental items would be auctioned.

Finally the "tangibles" would be sold to anybody with coupons left to buy them.

During experiment A, individuals would be allowed to have as many coupons in their account as they could acquire. In experiment B, individuals would be restricted to a figure equal to double the average number of coupons per person in circulation.

A number of different variables can then be observed. These would include the total amount of work done on the bikes (actually measurable in kW), prices of rental for the bikes, television rooms and sleeping rooms, and of course the total distribution of wealth in the community. It would also be possible to observe and interview the various participants during the duration of the experiment and gauge their levels of enjoyment and/or frustration.

It is the personal belief of the author, that in the first experiment, a small number of enterprising individuals would drive up the rental price of the bikes and become rich enough to buy a large proportion of the tangibles. In the second experiment it is believed that the bike rental price would remain low and the distribution of tangibles would be much more even. Whether these beliefs are correct or not would be for the experiment to show.

Clearly human beings are not as easy as molecules to control, and there are potential problems with the above experiment. One obvious problem being deliberate collusion; very difficult to avoid with the small numbers involved. The second obvious problem, especially with the second experiment; is the creation of "off the record" dealing and a black market using an illegal currency trading cigarettes or similar. Failure in this manner would in itself be useful research, it would allow estimates to be made of the real life workability of such schemes, and whether human beings can be "engineered" by the application of econodynamic theory.

The above formally proposes a controlled experiment that predicts in advance a symmetrical income distribution for experiment B; a previously unobserved economic phenomenon. Prediction of this sort is unusual in the field of economic research and potentially very persuasive.

If successful, and the different expected distributions were produced in the two parts of the experiment this would not only support the applicability of the particular ideas in this paper. It would also be a powerful indication of the general applicability of econodynamics and econophysics.

## 5. Conclusions

Econophysics is a relatively new science; while dramatic intellectual insights have been made, progress to date has largely been observational, with explanations being given for existing phenomena.

In this paper an engineering approach has been taken that uses assumed knowledge of the underlying mechanisms of wealth / income distributions (the "econodynamics" of Solomon and Richmond) to propose possible effective ways of changing economic systems.

It is highly likely that these initial ideas are far too simplistic to be practical in the forms described above. It is also the case that the decisions to make such changes would be essentially political.

It is hoped however that the ideas above show a possible line of enquiry that could prove more practical in redistributing income in the long term than current policies of transfers of taxation to welfare.

Finally, in line with good scientific practice, possible experimental verification is offered for the authors ideas.

## 7. Appendix: Hamburger Wars

The following section attempts to describe in a simplistic manner how market forces would drive an alternative welfare system which uses the closed systems described above in section 3 above.

It discusses two redundant labourers at the bottom of society. Call them Jill and Ken. For reasons of inexperience, outdated skills, disability, indolence etc; we will assume that they are two-thirds the use of a normal labourer. That is, to get £10 of useful labour out of them we have to pay them £15 (2/3rds of £15=£10). Which is to say, they have "priced themselves out of the market".

Let us assume that a normal worker does 40 hours work for £3/hour so receiving a wage of £120. Then if Jill or Ken work a 40 hour week they only produce a value of £80 (2/3rds of £120=£80) or £2/hour (£80/40hours=£2/hour). But they still expect to be paid £3 for each hour worked.

At the present time Jill and Ken are sat at home getting £40 dole each week from the government.

We are also going to have two potential employers of these people. Both are reputable fast food merchants, MacRonalds (MR) and their rivals Burger Queen (BQ). Our plan is to persuade these two companies to employ Jill and Ken by paying subsidies to the companies. However, we are not going to pay them the subsidies directly. We are going to use the free market to make MacRonalds and Burger Queen compete against each other for the subsidies. We are also going to force Jill and Ken to compete with each other. We are going to issue Jill and Ken with a weekly "Allowance" for 120 "Coupons" call this allowance: A120. For every coupon that J or K collect the government will pay them £1, up to the limit of their allowance. So if Ken collects fifty coupons (C50) and hands them in at the post office the person on the counter will exchange the coupons for £50 cash.

Before we go any further I would like to make clear the distinction between allowances and coupons. An allowance of A120 is a piece of paper given to a person by the government allowing that person to cash in a maximum of 120 coupons in a certain week. To get the full £120 the person has to also collect 120 coupons. The coupons are to all intents and purposes a new currency which can be swapped between anybody who is interested in exchange for goods or much more importantly, labour. The point is that in the long run the coupons are only really useful to people who have allowances, because these people are the ones that can use them to get real money from the government.

In the meantime the government will sell the coupons, to anybody who wants to buy them, on the open market. Maybe also at the post office. However each week the government sells only C200. Even though J and K's total allowance is A240 (A120x2=A240).

Before we start first note that the government is paying out a total of £80 dole to Jill and Ken (£40x2=£80).

Week 1.

Jill goes to the Post Office and buys C120 and offers only £5 for them. She does not buy any more because her allowance is only A120. She then cashes in her C120 against her A120 and receives £120. She is now £115 up on the deal (£120-£5=£115). The price of coupons is now 4.2 pence each (£5/120Coupons=£0.042=4.2pence/coupon).

Ken comes in later and gets the remaining C80 for £3-33 (80Coupons x 4.2p/coupon =333p=£3-33). He cashes them in against his A120 and receives £80. His net earnings are then £76-67 (£80-£3-33=£76-67).

Week 2.

Ken gets up a bit earlier and offers £10 for C120 which he cashes in for £120, making a profit of £110 (£120-£10=£110). The price of a coupon has gone up to 8.3p (£10/120Coupons=8.3p/coupon). This week Jill gets only the remaining C80, at a cost £6-67 (C80x8.3p/C=£6-67), and so gets £73-33 in her hand (£80-£6-67=£73-33).

Week 3.

Jill pays £20 for C120 and nets £100. Coupons now cost 16.7p each (20/120). Ken pays £13-36 (80x16.7) for C80 and nets £66-64 (80-13.36).

Week 4.

Ken pays £40 for C120 and nets £80. The market price of coupons is 33 pence each (40/120). Jill pays £26-40 (80x33) for C80 and nets £53-60 (80-26.4).

Week 5.

Jill pays £60 for C120 at 50p/coupon, and nets £60. Ken pays £40 for C80 and nets £40.

At this point the competition stops as the loser can get £40 just as easily by signing on the dole. The government is paying out a total of £200 (£120 to Jill + £80 to Ken), but is receiving payments of £100 (£60 from Jill + £40 from Ken), so is only paying out £100 net. This is only twenty pounds more than the government was paying in dole at the beginning of the exercise. In fact with a lot of people in competition, beggar my neighbour tactics would tend to reduce the payouts back to the same level as the dole.

So one day Ken is complaining to his friend about the Governments new scheme. It had started so well, but now he might as well be on the dole.

His friend happens to be the manager of the local MacRonalds and she suggests a new deal for him. If he is willing to work for 40 hours at MR she will give him 120 coupons.

Week 8.

On Monday morning Ken reports for work at MacRonalds. Meanwhile the MR manager goes to the post office and buys 120 coupons at 50p each, which is the price they were selling at last week (Week 7). That costs MR £60 (C120x50p/C=£60). At the end of the week Ken has completed 40 hours of work and MR give the C120 to him. Ken cashes these in for £120 at the Post Office and this week he has £120 in his hand for 40 hours work. Ken has received 3 coupons (and so £3) for each hour of his work and so he does not feel cheated. (Remember we agreed at the beginning of the chapter that Ken wants to be paid at the market rate of £3/hour even though he is 2/3rds efficient and only produces work worth £2/hour. We are going to call this L2/h.) During Week 8, MR has received labour of a value of eighty pounds (40hours x L2/h=L80), but they only paid £60 for the coupons, so they have made a profit of twenty pounds. Finally, this week Jill is forced to buy C80 at 50p each, and only has £40 in her pocket.

Week 9.

Having closely watched MacRonalds last week, this week Burger Queen step in and offer Jill a similar deal. They get to the post office before MR and buy C120 at 50p each. They give these coupons to Jill in exchange for 40 hours of labour. She cashes them in for £120. This week it is BQ that pay £60 for L80 and they make £20 profit.

Having arrived late at the Post Office MR can only buy the 80 coupons that BQ have left behind. Because they only have 2/3rds of the coupons that Ken wants, they agree he should only work 2/3rds of the hours. So MR give the C80 to Ken in exchange for 27 hours labour. MR paid £40 for the coupons (C80x50p/C=£40). They received L54 from Ken (27hours xL2/h=L54) and so make £14 profit (L54-£40=£14).

Week 10. Now the competition starts in earnest. Having been beaten last week, MR now up the ante and bid 65p per coupon (last weeks price was 50p/C). They buy C120 which now costs them £78 (C120x65p/C=£78). They give the coupons to Ken, again in exchange for 40 hours work. Again he still provides MR with L80. MR are still making a profit but this time only £2 (L80-£78=£2).

This week BQ can only buy C80 also at 65p/Coupon, this costs them £52 (C80x65p/C=£52). Jill only works 27 hours this week, producing L54. BQ also make a profit of only £2 (L54-£52=£2).

At this point lets check the overall figures:

The government is paying Ken £120 when he cashes in his coupons, they are also paying Jill £80 for her coupons. This is a total of £200 the government is paying out. MR is paying the government £78 to buy coupons for Ken, BQ are paying £52 to buy coupons for Jill. This is a total of £130 the government is receiving. The government is only paying out a net sum of £70 (£200-£130). This is £10 Less than the £80 dole the government was paying out at the beginning of the exercise.

We can also note the following:

i) Ken and Jill both have jobs.

ii) Ken is taking home three times what he was getting on the dole, Jill merely double.

iii) The two fast food companies are both making a modest profit.

iv) The economy as a whole is receiving an input of labour to a value of £133 (L80 + L53).

Profit for the two companies becomes zero at a price of 66.7 pence per coupon. This is because our two employees are only 66.7% (two-thirds) efficient. We assumed this at the beginning of the exercise. What happens if we change these assumptions?

During week 10, suppose BQ manage to increase Jill's efficiency from two-thirds, to three-quarters, of that of a typical worker's. They might do this by training her on a new burger flipping machine or simply by offering to give her more hours work, and so more coupons, if she will work that little bit harder. So now Jill produces £2.25 worth of labour for each hour she works instead of only £2.00. (Typical worker: L3/hour, Jill is now 3/4 as efficient; L3/h x 3/4= L2.25/h.)

Week 11.

BQ buy C120 still at a price of 65p per coupon, this costs them £78. They give them to Jill for 40 hours labour, which now produces a value of L90 (40h x L2.25h=L90). BQ are now making a profit of £12 (L90-£78=£12). This week Ken loses out as MR can only buy C80 for him. This week he only works 27 hours. He is still two thirds efficient so he produces L54. MR paid £52 for the C80 so they still make a small profit of £2.

There are two important things to note here. Firstly the shortage of coupons makes it worthwhile for both employers and employees to compete. Because BQ are making a bigger profit out of Jill they can afford to fight harder to get more coupons for her. So she also benefits. It is worthwhile for BQ to make Jill more efficient, it is also worthwhile for Jill to become more efficient.

Secondly although Ken loses this week, he is still doing quite well. He is earning twice what he would on the dole.

Looking at this more generally, we issued 200 credits which is about 80% of the total allowances available (A240). By doing this we have created a "closed system". In this market there is still competition; there are still winners and losers, but in this case the laws of statistics ensure that most people are at the top of the scale.

Suppose now we had 10 people all with an allowance of A120 giving a total of A1200. The government would issue C100 per person or a total of C1000. Now in theory it is possible for eight people to earn C120, one person to get only C40 and the last person to get nothing. But statistically this is almost impossible.

Below are some random results run on a simple computer model. The model was run 10 times to give more results. Each time there were 10 people with an average wealth of C100 and a maximum wealth of A120.

| wealth range | 0-9. | 10-19. | 20-29. | 30-39. | 40-49. | 50-59. | 60-69. |
|---|---|---|---|---|---|---|---|
| No of people | 0 | 0 | 1 | 0 | 0 | 2 | 5 |

| wealth range | 70-79. | 80-89. | 90-99. | 100-109. | 110-119. | 120. | Total |
|---|---|---|---|---|---|---|---|
| No of people | 4 | 8 | 17 | 29 | 30 | 4 | 100 |

Taking these results as typical, then 63% of the people will get more than 100 credits. Only 29% will get between C100 and C70, and fewer than 8% will earn less than C70. Only 1% has got less than the original dole level of C40.

If we pluck some very rough figures out of the air. Assume the average wealth in our society is £200 per week, then the poverty level is defined as half the average or £100, we can define extreme poverty as less than one third of the average ie below £70. From our discussions above, in a present day, open system based on a Boltzmann distribution, we would expect about 15% of the population to be below the poverty level and, of these, maybe a third (5-6% of the total population) will be in extreme poverty.

We can now use our new benefit system to aid this 15%. Over 9% are removed from poverty (63%of15%=9% are now above C100). Slightly over 4% (29%of15%) are now between C100 and C70. And only slightly over 1% remain in extreme poverty (8%x15%). These are dramatic improvements.

If the system proves workable then the market price of the coupons will rise as the skills of the low paid increase and they became more efficient. As the market price of the coupons rises the government will be spending less and less money. Then the government will have the resources to extend the scheme. Suppose they multiply the benefits by a factor of 1.4. The average wage is still

£200, but the allowances are now A170 instead of A120. The number of coupons is now increased from an average of C100 to C140 so they are still 80% of the total allowances.

Now the number earning less than C100 is about 1% and the number earning less than C70 is less than half a per cent. In practical terms poverty has been abolished. This remaining 1% will be the genuinely needy, who are physically or mentally unable to compete. These can best be dealt with by old fashioned methods of welfare and social work.

A trial schemes could be set up in unemployment blackspots. Monthly allowances could be issued with dole cheques, the size of the allowance would depend on the previous year's earnings, this is easily calculated from tax information. The lower the previous year's income the higher the allowance. People who get their hands on vouchers would then cash them in at the Post Office against their allowances. This information would allow the Dole office to calculate the next months dole and allowance. The vouchers themselves could simply be a paper currency sold through Post offices and banks (they would most certainly need to be photocopy proof). The total amount of coupons released would be a fixed fraction of the allowances issued, say 80%. Some initial publicity would be needed to encourage employers to use the coupons, but once people started making profits the scheme would quickly grow.

If the scheme proved workable, then the whole benefits and allowance scheme could be integrated into existing computerised tax systems. People and companies could even have bank accounts in coupons. In the long run the government would not even have to count the amount of allowances issued, they would simply sell coupons on the open market and monitor the price. If the price goes down, reduce the amount of coupons in circulation, if the price goes up increase the amount in circulation.

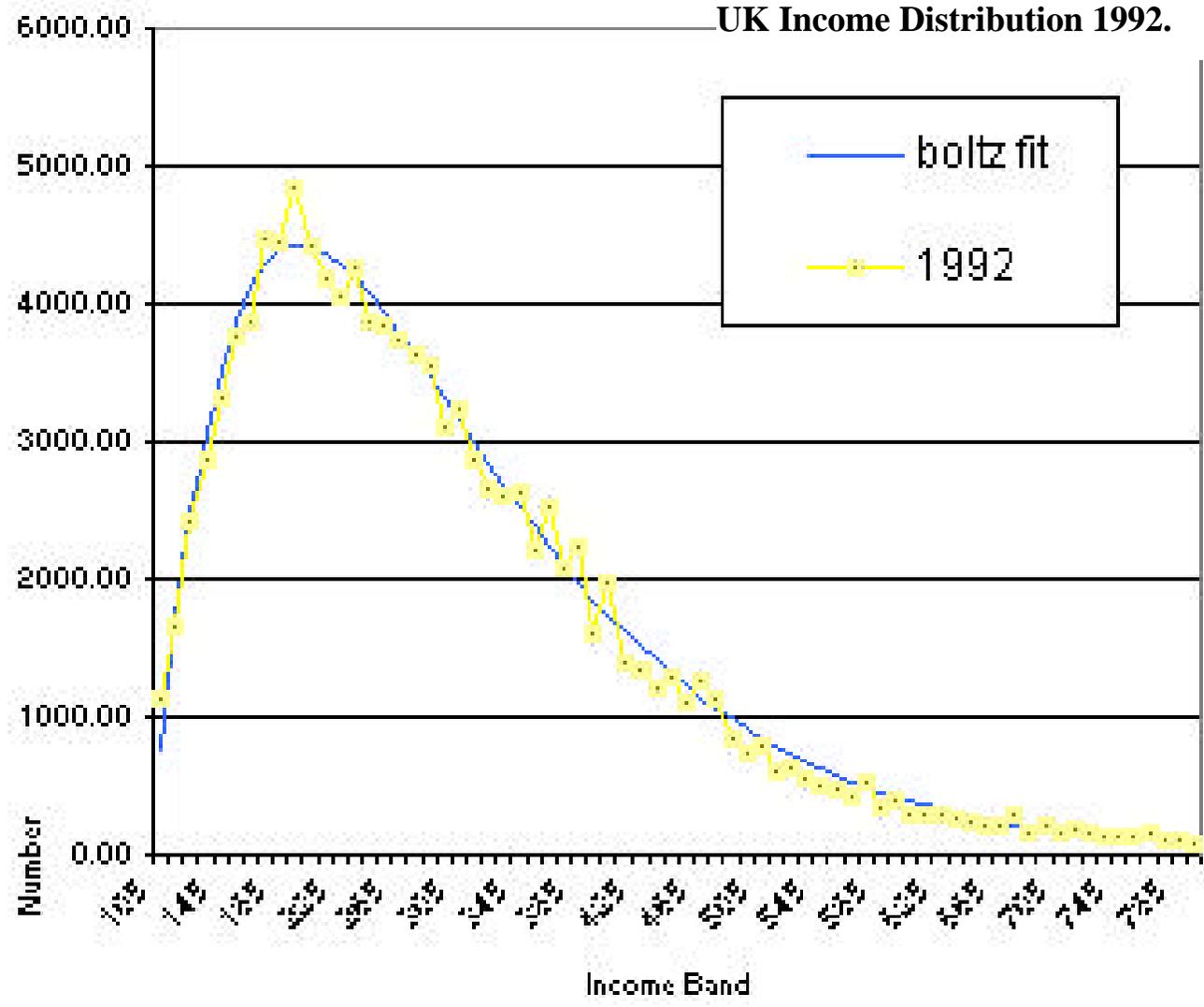

Figure 1
UK Income Distribution 1992.

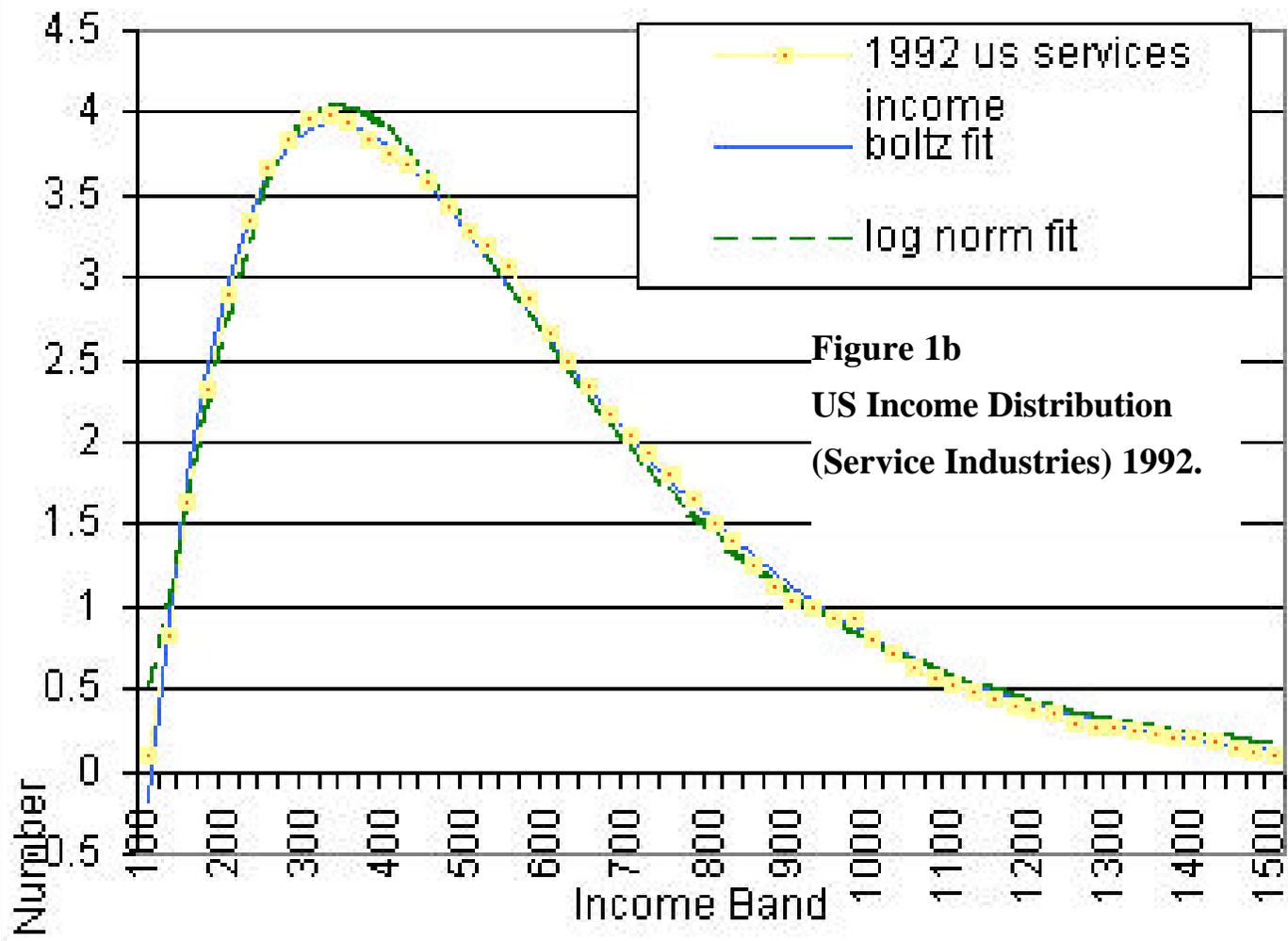

**Figure 1b**
**US Income Distribution**
**(Service Industries) 1992.**

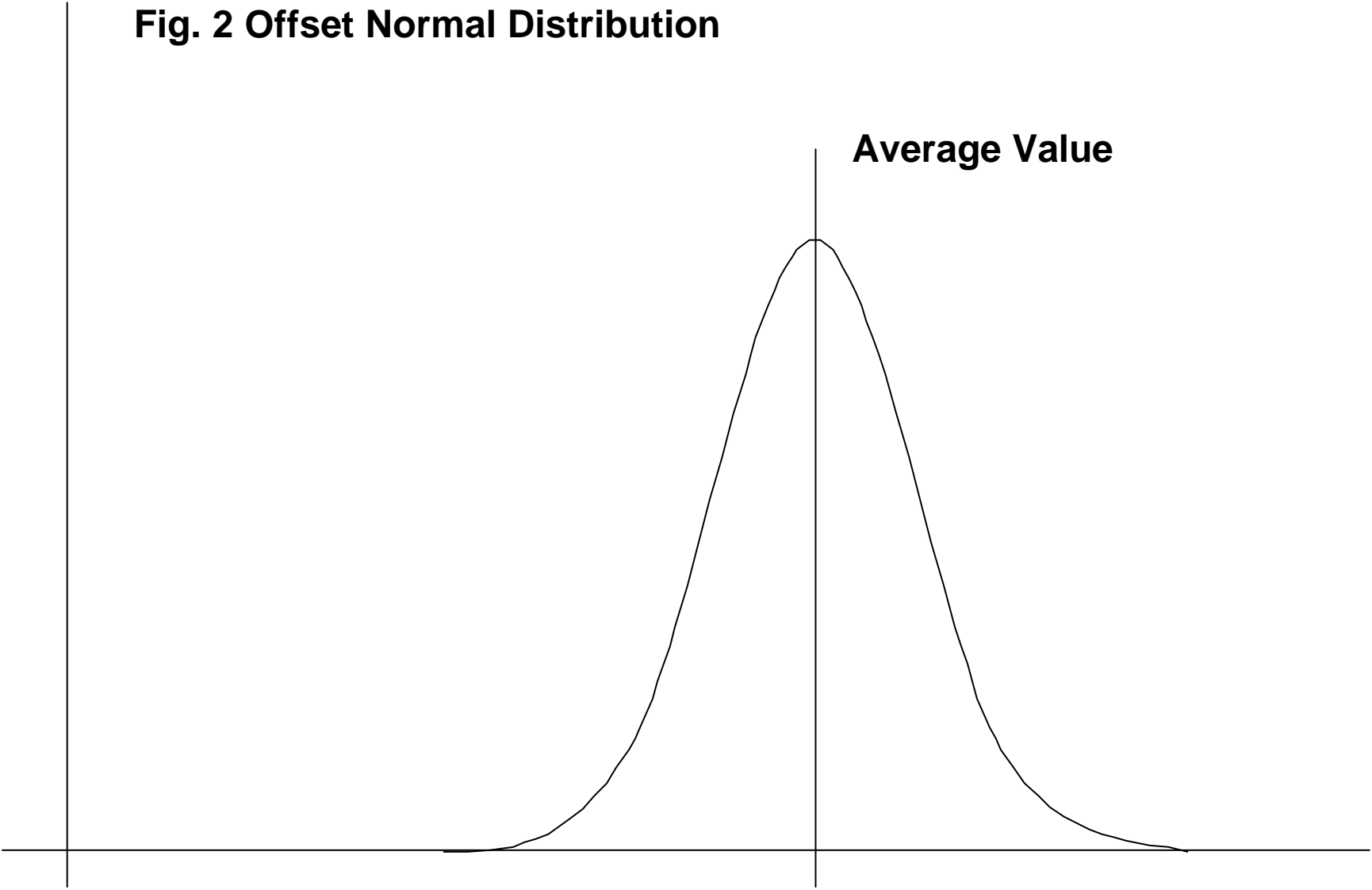

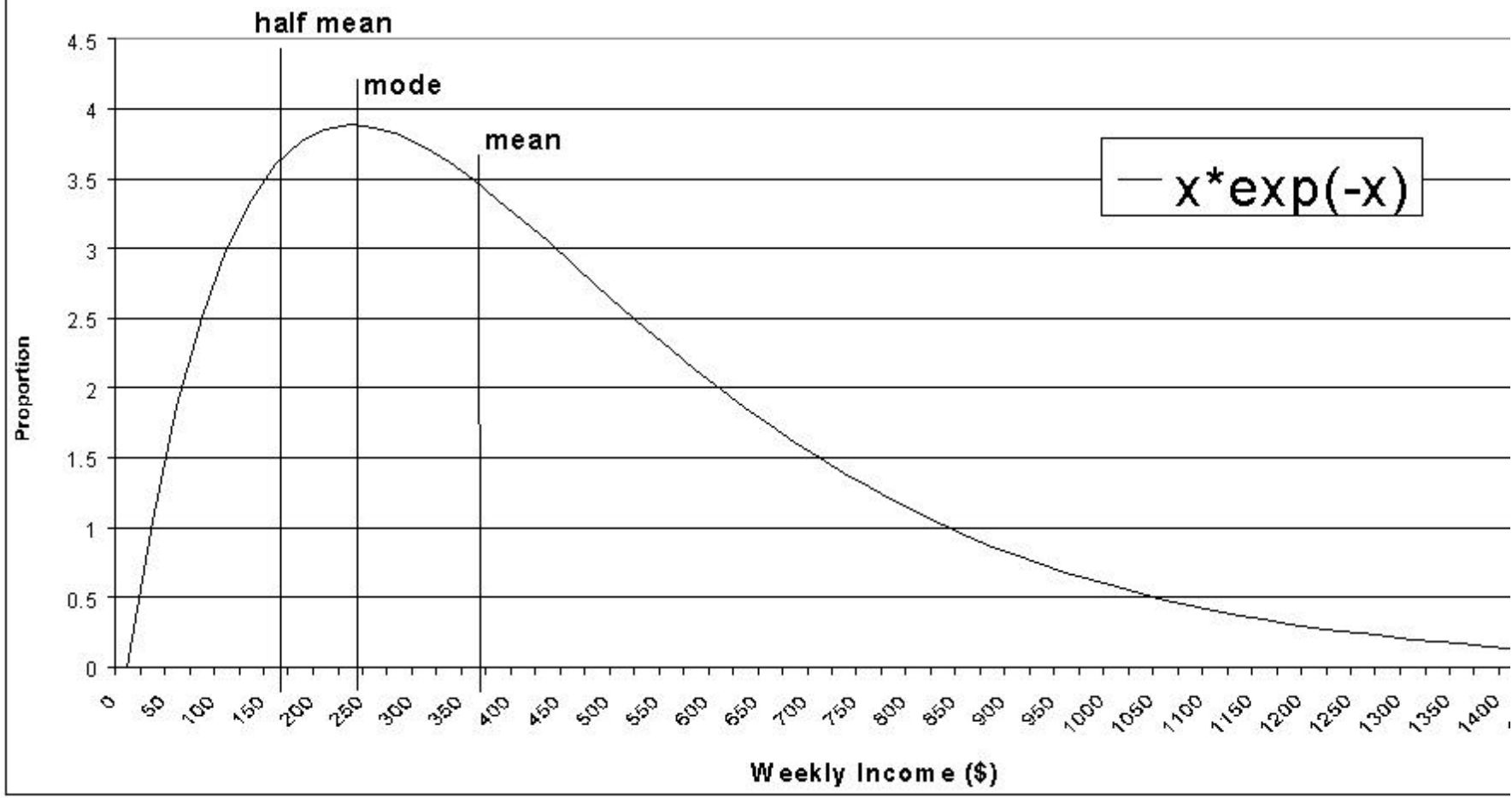
Fig.3 Boltzmann distribution

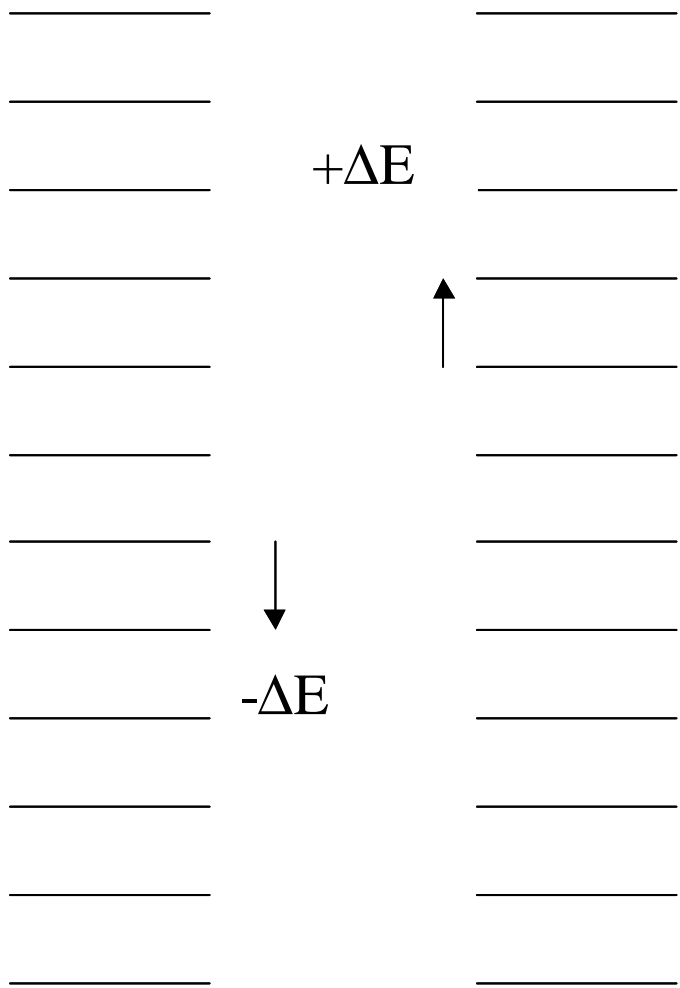 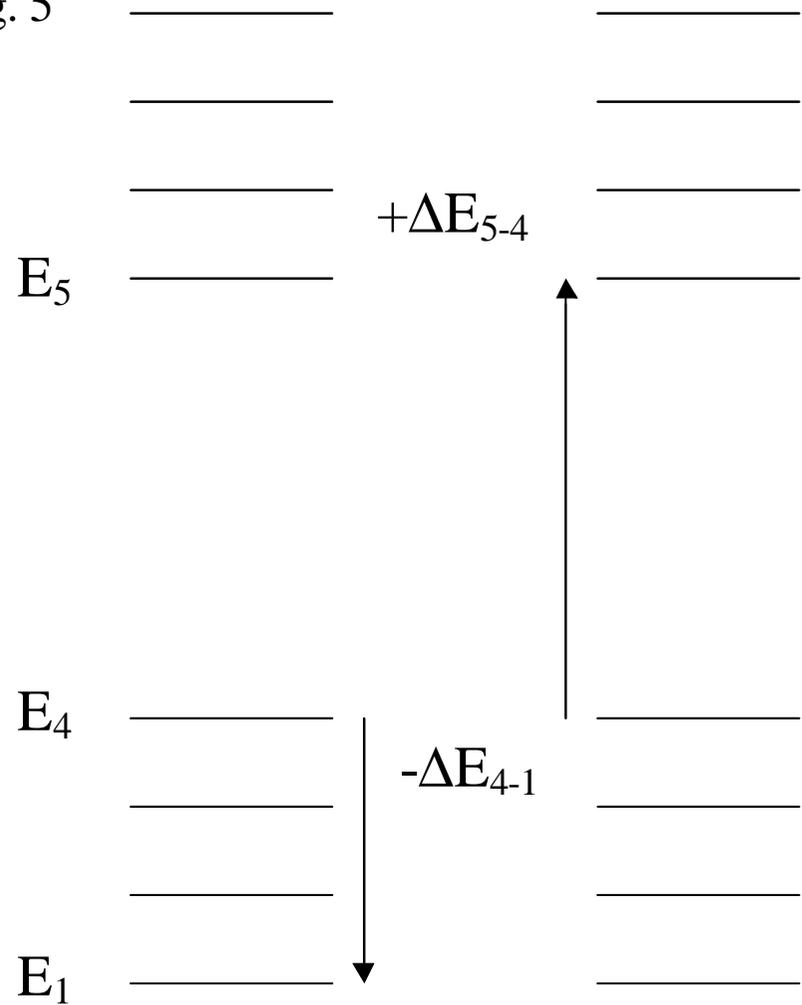

Fig. 4 Non lasing material

Energy exchange to upper levels possible

Fig. 5 Lasing material

Energy exchange to upper levels not possible

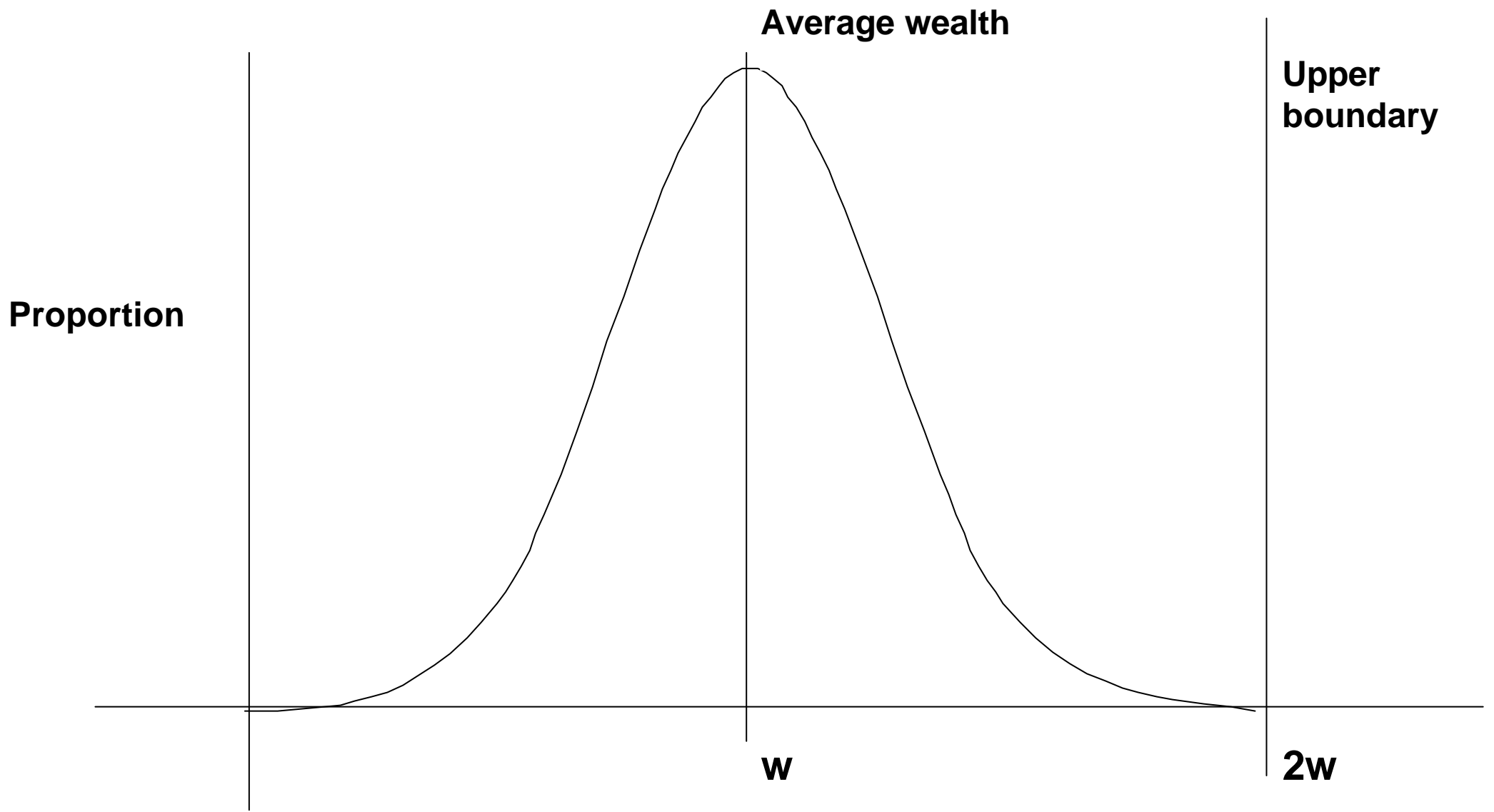

Fig. 6

**Fig. 7**

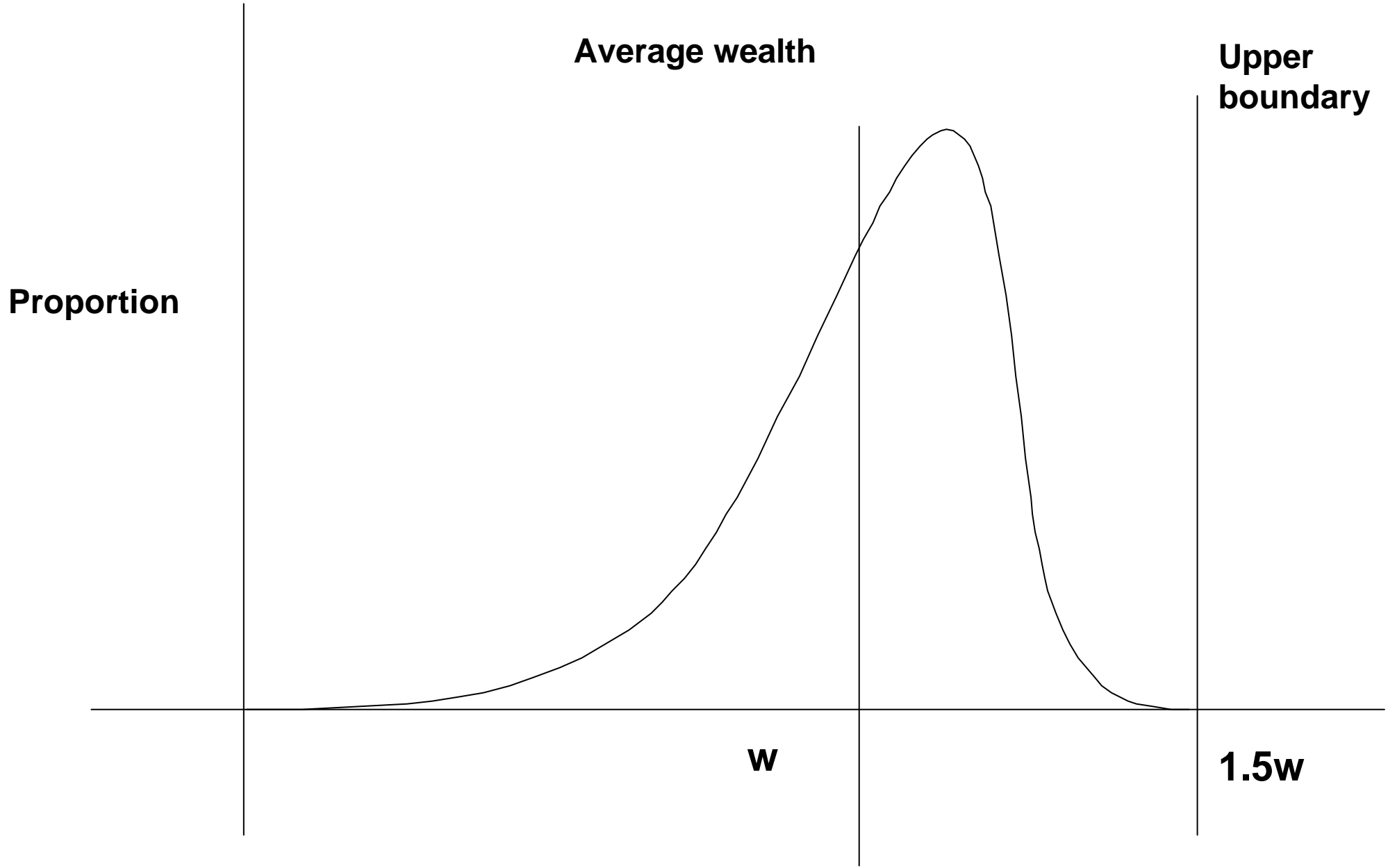